\documentclass[sigconf,screen]{acmart}
\AtBeginDocument{%
  \providecommand\BibTeX{{%
    \normalfont B\kern-0.5em{\scshape i\kern-0.25em b}\kern-0.8em\TeX}}}

\copyrightyear{2022}
\acmYear{2022}
\setcopyright{acmlicensed}
\acmConference[A-TEST '22]{Proceedings of the 13th International Workshop on Automating Test Case Design, Selection and Evaluation}{November 17--18, 2022}{Singapore, Singapore}
\acmBooktitle{Proceedings of the 13th International Workshop on Automating Test Case Design, Selection and Evaluation (A-TEST '22), November 17--18, 2022, Singapore, Singapore}
\acmPrice{15.00}
\acmDOI{10.1145/3548659.3561309}
\acmSubmissionID{fsews22atestmain-id4198-p}
\acmISBN{978-1-4503-9452-9/22/11}

\usepackage[normalem]{ulem}
\usepackage{soul}
\usepackage[utf8]{inputenc}
\usepackage{listings}
\usepackage{xcolor}
\usepackage{graphicx}
\usepackage{tikz}
\usepackage{url}
\usepackage{microtype} 
\usetikzlibrary{fit,automata,positioning,calc}
\usepackage{algorithmicx}
\usepackage{algorithm}
\usepackage[noend]{algpseudocode}

\newcommand{\Samira}[1]{{\color{brown}\small$\langle\langle$ #1 $\rangle\rangle$}}
\AtBeginDocument{%
  \providecommand\BibTeX{{%
    \normalfont B\kern-0.5em{\scshape i\kern-0.25em b}\kern-0.8em\TeX}}}

\AtBeginDocument{%
  \providecommand\BibTeX{{%
    \normalfont B\kern-0.5em{\scshape i\kern-0.25em b}\kern-0.8em\TeX}}}    
\newcommand{\MyNOTE}[1]{{\color{blue}\small$\langle\langle$ #1 $\rangle\rangle$}}

\newcommand{\HIDE}[1]{}
\begin{document}

\title{An Online Agent-Based Search Approach in Automated Computer Game Testing with Model Construction}

\author{Samira Shirzadehhajimahmood}
\affiliation{%
  \institution{Utrecht University}
  \orcid{0000-0002-5148-3685}
  \country{the Netherlands}}
\email{S.shirzadehhajimahmood@uu.nl}

\author{I. S. W. B. Prasetya}
\orcid{0000-0002-3421-4635}
\affiliation{%
  \institution{Utrecht University}
  \country{the Netherlands}}
\email{S.W.B.Prasetya@uu.nl}

\author{Frank Dignum}
\orcid{0000-0002-5103-8127}
\affiliation{%
  \institution{Umeå University}
  \country{Sweden}}
\email{fpmdignum@uu.nl}

\author{Mehdi Dastani}
\affiliation{%
  \institution{Utrecht University}
  \country{the Netherlands}}
\email{M.M.Dastani@uu.nl}

\renewcommand{\shortauthors}{Samira Shirzadehhajimahmood, I. S. W. B. Prasetya, Frank Dignum, Mehdi Dastani}

\begin{abstract}
The complexity of computer games is ever increasing. 
In this setup, guiding an automated test algorithm to find a solution to solve a testing task in a game's huge interaction space is very challenging.
Having a model of a system to automatically generate test cases would have a strong impact on the effectiveness and efficiency of the algorithm. However, manually constructing a model turns out to be expensive and time-consuming. In this study, we propose an online agent-based search approach to solve common testing tasks when testing computer games that also constructs a model of the system on-the-fly based on the given task, which is then exploited to solve the task. To demonstrate the efficiency of our approach, a case study is conducted using a game called Lab Recruits.
\end{abstract}




\maketitle

\section{Introduction}
Recently, the computer games industry has seen the emergence of advanced 3D games. These are often complex software due to their high level interactivity and realism. There is already a large body of research in automated software testing, proposing various methods to decrease the manual effort.
However, game testing is more complex in comparison to more traditional software testing. In games, the search space is huge, with no obvious structure. 

In automated game testing, computer controlled player-characters (agents) are used to test various aspects of a game, e.g. to verify that a certain objective in a given game level is achievable and is in the correct state. It would benefit testers if testing tasks can be formulated abstractly.
We then rely on the agent to automatically execute such a task by {\em searching} for a 'solution': a right sequence of interactions that would bring the agent to the task objective, to subsequently verify the objective's state.
Shirzadehhajimahmood et al. showed that such a test is also robust (can cope better with development time changes) \cite{shirzadehhajimahmood2021using}, because the solution is searched dynamically, rather than manually prescribed. To make this works, the searching part is  crucial. However, it is also the harder part to automate, due to the huge interaction space,  navigability, various game rules (e.g. limited vision, players cannot see nor walk through a solid wall), and long and complex game scenarios.

 

In computer games, solving a testing task requires a specific sequence of actions to be taken; just randomly or greedily trying them out does not work.  In addition, games typically have elements that resist the player, e.g. obstacles and hazards. When trying to solve a task, an agent must also deal with these elements, which is non-trivial as it may involve searching certain game objects and controlling them e.g. to unblock some obstacles. 
Solving this by applying the usual search based testing algorithm, such as evolutionary \cite{yu2010introduction}, directly on the game under test is not a workable option due to excessive computation time.
Having a behavioral model of the system under test would help. Ferdous et al. applied model-based testing to automate the generation and the execution of test cases from an Extended Finite State Machine (EFSM) model \cite{ferdous2021search}. However, constructing a model has to be done manually, and hence costly. A major challenge faced by the game industry is the lack of automated approaches for generating a model of the system under test (SUT).    

In this paper, we propose an online agent-based search approach to do automated testing on modern computer games. Being an {\em on-line} search approach, it does {\em not} require a full pre-constructed model of the game under the test. Rather, given a model (EFSM) that is only {\em partially specified} to capture only general properties of the game, the remaining part of the model is constructed {\em on the fly} during the search and exploited to aid the search process.
The approach is implemented on top of the {\em agent-based} testing framework iv4XR \cite{prasetya2020aplib}.
Using agents is an appropriate approach to deal with the high-level interactivity of computer games \cite{prasetya2020aplib,shirzadehhajimahmood2021using} thanks to agents' inherent reactive programming model.
We also benefit from other agents' related features such as goal-oriented behavior and the possibility to do autonomous planning to make the programming of test automation more abstract.





{\bfseries Paper structure.} This paper is organized as follows. Section \ref{sec.problem-setup} describes the setup of our approach. Section \ref{sec.modeling-game} discusses the kind of models that our algorithm constructs.
Section \ref{sec.online-agent-based search approach} presents our online agent-based search approach. Section \ref{sec.prolog} describes how to construct the aforementioned model.
Section \ref{sec.implementation} discusses the agent-based implementation that we used.
Section \ref{sec.experiment} discusses experiments we conducted to asses the effectiveness of our approach. Section \ref{sec.relatedwork} and \ref{sec.concl} cover related and future work, respectively.

\section{Problem Setup} \label{sec.problem-setup}

We assume an agent-based setup, e.g. a la iv4XR \cite{prasetya2020aplib}, where a test agent is available to take the role of the player to control the game. 
We can abstractly treat a game as a structure:
\begin{equation} \label{eq.game-structure}
Game \ = \ (Nav,O,L)
\end{equation}
%
where $Nav$ is a structure describing the navigable terrain of the game world \cite{millington2019AI}, $O$ is a set of game objects, and $L$ is a set of  actions available to the test agent. 
Game objects have properties such as their positions, and being interactable or hazardous.
The agent also has its own properties, such as its position, and what it currently sees. 
Objects such as doors are called {\em blockers}; they can block access to other objects. 
Objects that can change the state of blockers are called {\em enablers}, for example switches and keys.
The overall game state, also called {\em configuration}, comprises of the properties of the objects and the agent.
The test agent is bound by typical game physics: it can only travel over navigable terrain ($Nav$), and it can only {\em observe} objects and parts of $Nav$ that it physically can see (e.g. it cannot see through a wall). So, initially $Nav$ usually contains only a part of the terrain where the agent starts.
Typically, primitive actions available to the agent are:
{\em moving} in any direction for a small distance, 
and {\em interacting} with an object $o$.
From these we assume the following high level actions can be constructed, 
which comprise the set $L$ in (\ref{eq.game-structure}); the construction was described e.g. in \cite{prasetya2020navigation}.
\begin{itemize}
    \item $navigateTo(o)$, to travel to the position of $o{\in}O$.
    This can be done by implementing a path finding algorithm such as A* \cite{hart1968formal, millington2019AI}, applied on $Nav$.
    %
    %
    \item $explore()$ incrementally explores the game world. 
    It stops when new terrain is sighted (and added to $Nav$).
    A graph-based exploration algorithm such as  \cite{prasetya2020navigation} can be used.
   \item $interact(o)$, to interact with $o$ as mentioned above.  
\end{itemize}

To test something the agent must be given a 'testing task'. An elementary type of tasks is to simply verify whether certain states of a game object $o$, 
characterized by a predicate $\phi_o$, are {\em reachable} from the game initial configuration $c_{init}$, and furthermore satisfy a certain correctness assertion $\phi$.
For example, $\phi_o$ can be "{\em a treasure chest becomes visible}", and $\psi$ asserts that the agent should by then collect enough game-points.  
\HIDE{
'Reachable' means that there exists a sequence of agent's actions
that leads to a configuration where $\phi_o$ is true. More precisely, $\phi_o$ is reachable if:
\begin{equation} \label{eq.testingSequence}
  \exists \delta,c.\ \  
   c_0 {=} c_{init} \And
   c_{0} \xrightarrow{\delta_0} c_1
   \ ... \ 
   c_{n{-}1} 
   \xrightarrow{\delta_{n{-}1}} c_n
   \And c_n \models \phi_o
\end{equation}
where $\delta$ is a sequence of high level actions as mentioned before, and $c$ is a sequence of configurations.
Note that to check this the agent will have to find such a $\delta$, which is not trivial, since at the start the agent does not even know how to unblock blockers. It is also not possible to know an action's resulting configuration without executing it first on the actual game under test.
}
%
%
%
%
\HIDE{Often, we also want to verify that the reachable part of $\phi_o$ also satisfies another predicate $\psi$; so, whether $c_n$ above also satisfies $c_n \models \psi$.} 
Abstractly, this can be formulated as:
\begin{equation} \label{eq.testingtask}
    \underbrace{\phi_o}_{\rm situation\ required\  to\ be\  reachable} \Rightarrow \underbrace{\psi}_{\rm assertion}
\end{equation}
\HIDE{The form is simple, but still represents many useful testing tasks. 
The term "assertion" refers to a property in a game world that we want to assert.
For example, $\phi_o$ can specify that a treasure becomes visible to the agent, and $\psi$ asserts that the agent should by then collect enough game-points.  
}
Complex tasks can be built by composing elementary tasks.

The problem to solve is to automatically perform a testing task, given only a description as above. Note that this is a {\em search problem}:
\HIDE{We treat a testing task, more specifically the $\phi$-part, as a goal that a test agent wants to automatically achieve/solve (and thus providing test automation). To do this} the executing test agent needs to find a right series of actions that reaches a state satisfying $\phi_o$, {\em while respecting the game rules}.
This search is far from trivial. Checking the assertion part is usually easy.
Our automated 
approach 
will consist of these two key elements: 



{\bf On-the-fly Model.} The search would be more effective if we have a model as in model-based testing. However, since we do not actually have a model, our search algorithm builds one on-the-fly, and exploits it to help the search. 
%
%
We use Extended Finite State Machine (EFSM) as the model, with a twist 
%
so that the EFSM also captures physical navigability over $Nav$. 

{\bf Online search.} The proposed search approach, presented in Section \ref{sec.online-agent-based search approach}, is an online search, where the agent directly explores the game under test. The benefit is that the agent can access accurate state information from the game.  
A key element of the approach is dealing with obstacles, which can have a great impact towards solving the reachability part (the $\phi$-part) of a testing task.

\section{Hybrid Models of Games}\label{sec.modeling-game}

\begin{figure}[h]
\centering
\includegraphics[width=\linewidth]{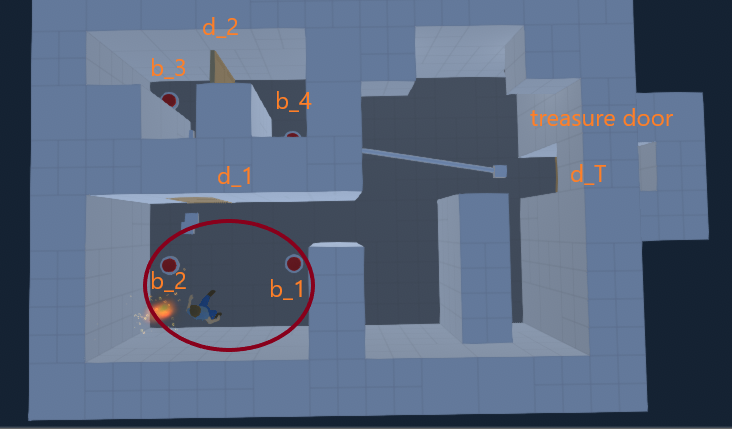}
\caption{A screenshot of a level in a game called Lab Recruits. The level's objective is to open the treasure door.}
\label{fig.LR}
\vspace{-0.5mm}
\end{figure}

As mentioned, our search algorithm constructs a model as it goes. 
More precisely, an Extended Finite State Machine (EFSM) model will be constructed. 
EFSM is expressive and commonly used for modelling software systems \cite{alagar2011extended}.
\HIDE{
 In this section, we introduce the structure we use to model the play behavior of a game.
 The model is constructed simultaneously during the game testing and exploited. The detail of how it will be constructed is explained later in section \ref{sec.prolog}. 
}
This model should capture not only the logic of a game, but also relevant physical aspects of the world. This  poses an additional challenge. Consider a simple 'game level' shown in Fig. \ref{fig.LR}, taken from a maze-like 3D game called Lab Recruits \footnote{https://github.com/iv4XR-project/labrecruits}. 
To interact with an in-game button, e.g. $b_4$, the player should be close enough to it, which means the button should also be reachable.
So, when modelling a transition between states, in addition to considering what it does, the transition must be {\em physically} possible in the game world as well. 
Since the standard use of EFSM does not capture physical navigability,
we define a 'hybrid' variation of EFSM that also captures this. 
Also, games often have a concept of 'zone', so we add this as well. A zone is an 'enclosed' part of $Nav$ where the player can travel freely. Traveling to another zone has to pass through an open blocker, such as a door, that connects zones, or unblock it first, if it is blocked.

\HIDE{Before we introduce the structure of our EFSM, first note that in EFSM we have 'state' and 'configuration'. A configuration describes an actual/concrete state of an EFSM, whereas 'state' refers to some chosen abstraction of concrete states.}
Deviating from \cite{alagar2011extended}, we will represent our EFSM by a tuple: 
\begin{equation}
M \ = \ (
   \underbrace{S,\ \ T,\ \ \Sigma,\ \ P}_{to\ be\  constructed\ \ },\underbrace{{\alpha},\ \  c_0}_{given\ by\  developer})
\end{equation}

The last two components should be provided by the game developer; the rest is learned/constructed on the fly.
$S$ and $T \subseteq S\times L \times S$
describe the states and transitions of $M$, as known to the agent so far. 
$L$ is the set of available actions listed in Section \ref{sec.problem-setup}.
Members of $S$ are also members of $O$ in (\ref{eq.game-structure}).
Being in the state $o{\in} S$ is to be  interpreted as: {\em the agent is currently at the game object $o$'s location}. As mentioned, objects have their own properties; their values define the EFSM's extended state.
Transitions in $T$ represent physical travel on $Nav$: when two different states are connected by a transition, it means that there is a path in $Nav$ between the two game objects represented by the states, that does not go through a blocker in between.
$M$'s other components:

\begin{itemize}
   
    \item $\Sigma$ is a set of aforementioned 'zones' in the game. 
    \HIDE{A zone $R$ is a subset of $S$ whose members are freely reachable from one another. More precisely, a path of transitions in $T$ should exist between any two members of $R$, and furthermore any path that crosses two zones must pass a blocker $o\in S$. Blockers connect two adjacent zones; and as such every blocker will be a member of exactly two zones.
    }
    
    \item $P \subseteq S \times S$; when $(i,o) \in P$ it means that the agent has learned that interacting with $i$ affects the object $o$. 

    \item $\alpha$ is a function that models the effect of $interact(i)$ on the objects in $O$, given the knowledge
    in $P$.
    
    \HIDE{
    ${\alpha} : {\mathcal L} {\rightarrow} ({\mathcal D} {\rightarrow} {\mathcal D})$ models the effect of an action on the objects. The action $navigateTo(o)$ does not change objects' properties. So, ${\alpha}("navigate(o)") = Id$, where $Id$ is the identity function. 
   For other types of actions, the developer has to provide their model.
   }

    \item $c_0$ is the initial 'configuration' of the system when it starts. 
    A {\em configuration} describes a concrete state of $M$ (as opposed to 'abstract'
    states $S$). It is represented by a pair $(s,D)$ where $s\in S$ (describing
    the agent's current physical location) and $D$ is a vector of all objects'
    properties in $S$.

\end{itemize}

\HIDE{represented as a pair $c = (\sigma,D)$ where $\sigma$ is a sequence over $S$ representing the history of the system states until now
and $D{\in}\mathcal D$. The last state in $\sigma$ is $M$'s current state, denoted by $current(\sigma)$. 
The current values of the properties of the entities in $S$ are as described by $D$. }

\begin{figure}[h]
\begin{tikzpicture}[
    state/.style={
      rectangle,
      rounded corners,
      draw=black,
      font=\sf\scriptsize,
      inner sep=3,
      fill=gray!10,
      text centered},
    every edge/.append style={font=\sf\scriptsize}
]
\node[state] (B1) at (-0.8,0)    {$b_1 (R_1$)};
\node[state] (B2) at (-0.8,-2)   {$b_2 (R_1)$};
\node[state] (B3) at (4.8,0)    {$b_3 (R_2)$};
\node[state] (B4) at (4.8,-3)    {$b_4 (R_3)$};
\node[state] (D1) at (2,-1)    {$d_1 (R_1,R_2)$};
\node[state] (D2) at (4.8,-2)    {$d_2 (R_2,R_3)$};
\node[state] (DT) at (-0.8,-3)    {$d_T (R_1,R4)$};
\draw  
    (B1) edge[<->,pos=.5,sloped,below] node{$navigateTo$} (D1)
    (B1) edge[<->,pos=.5,sloped,below] node{$navigateTo$} (B2)
    (D2) edge[<->,pos=0.5,sloped,below] node{$navigateTo$} (D1)
   (B3) edge[<->,left] node{$navigateTo$} (D2)
    (B2) edge[<->,right] node{$navigateTo$} (DT) 
   (B2) edge[->,loop left, looseness=10, below] node{$interact \mapsto \{ d_1\}$ \ \ }
   (B2) edge[<->,pos=.5,sloped,below ] node{$navigateTo$} (D1)
   (B3) edge[->,loop left, looseness=6,  left]node{$interact \mapsto \{d_1,d_2\}$}
 (B3) edge[<->,below left,pos=0,sloped] node{$navigateTo$} (D1)
  (B4) edge[->,loop left,looseness=6, left] node{$interact \mapsto \{d_T\}$ } 
  (B4) edge[<->, left] node{$navigateTo$} (D2)
   ;
\end{tikzpicture}
{\small
\caption{An EFSM model of the level shown in Fig.\ref{fig.LR} 
The notation e.g. 
$b_1 (R_1)$
in a state means that the state represents the object $b_1$, and furthermore $b_1$ is in zone $R_1$.
The $P$-component of the model is described by an extra annotation e.g. $\mapsto \{ d_1 \}$ 
on $interact$ transitions. E.g. on $b_2$ it means that $(b_2,d_1) \in P$. So, toggling $b_2$ affects $d_1$.
}}
\label{fig.EFSM}
\vspace{-5mm}
\end{figure}
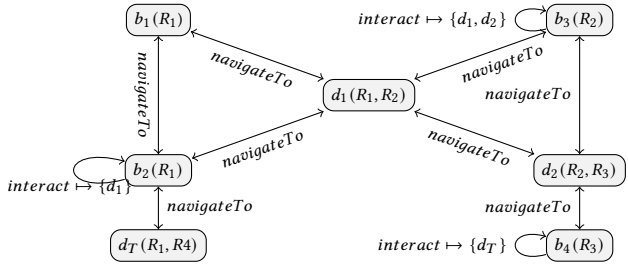

An example of as model is shown in Fig. \ref{fig.EFSM}.
The search algorithm in Section \ref{sec.online-agent-based search approach} does not need to do on-model execution; it relies only on the knowledge built in the first four components of $M$. 
However, we want to note that the constructed $M$ can be given to an off-line model based testing (MBT) algorithm such as in \cite{ferdous2021search} for generating test sequences. For this, on-model execution is needed. Off-line approaches are however outside this paper's scope.

\section{ Online Agent-based Search 
} \label{sec.online-agent-based search approach}

In this section, we provide the details of our automated online search algorithm to solve testing tasks. The algorithm takes three parameters shown in line \ref{search.alg.header} in Algorithm \ref{alg:onlineSearch}. The first, $\phi_o\Rightarrow\psi$, is a testing task as in (\ref{eq.testingtask}). The algorithm interacts with the game under tests, searching for a sequence of interactions that brings the game to a state satisfying $\phi_0$, and then it checks (line \ref{search.alg.assert}) if $\psi$ is satisfied. If it is, the test succeeds, and else a violation is concluded. 
\HIDE{For simplicity, the concrete game-state/configuration can be thought to have the same structure of an EFSM configuration as defined in Section \ref{sec.modeling-game}.}
Because the algorithm is implemented on an agent-based framework (Section  \ref{sec.implementation}), and in the agent  terminology $\phi_o$ is treated as a {\em goal},
\HIDE{we will also use this terminology. So,} the algorithm can also be thought as an algorithm for solving a goal.

The parameter $M = (S, T, \Sigma, P, \alpha, c_0)$ is an EFSM with the structure as in Section \ref{sec.modeling-game}, intended to model the game under test. The last parameter
$Nav$ is the navigable terrain of the game world mentioned in Section \ref{sec.problem-setup}. 
The $(S,T,\Sigma,P)$ part of $M$ is treated as a state-graph describing the game world; it will be denoted by $M_{stategraph}$. The sets of interactables and blockers in $S$ will be denoted by $I$ and $B$; so, $I\cup B \subseteq S$. Note that $M_{stategraph}$ and $Nav$ are
{\em initially empty}
As the algorithm  proceeds; these components will be incrementally built based on what the agent observes within its visibility range. 

\HIDE{
Imagine a testing task to verify if a state satisfying $\phi_o$ for an object $o$ is reachable and that this state satisfies some predicate $\psi$. 
The sequence of actions that leads to $\phi_o$ is not known upfront. The algorithm ${onlineSearch}(\phi_o\Rightarrow\psi,M, Nav)$ performs an online search to find such a sequence, and since it is online, we can note that searching the right sequence will also, as a side effect, take the agent to $o$, if it is found. Note that the agent visibility range is limited. The algorithm incorporates some heuristics/policy to guide the search, which are outlined below.
}

The algorithm actually performs a two levels search, though here we will focus on its higher level part.
The lower level is used to find a path to guide the agent to physically travel over walkable regions ($Nav$) of the game world.
In the setup defined in Section \ref{sec.problem-setup}, this functionality is encapsulated within the procedure $navigateTo$. The upper level of the search is used to abstractly search at the game-objects level; Algorithm \ref{alg:onlineSearch} is formulated at this level.
It incorporates some heuristics/policy to guide the search, which are outlined below.

\algblockdefx{ParStart}{ParEnd}{{\bf parallel}}{{\bf end parallel}} 
\algblockdefx{ParIfStart}{ParIfEnd}[1]{|| {\bf if} #1 {\bf then}}{{\bf \ \ end if}}
\begin{algorithm}
\small
\caption{Online Search}\label{alg:onlineSearch}
\begin{algorithmic}[1]
\Procedure{onlineSearch}{$\phi_o\Rightarrow\psi,M, Nav$} \label{search.alg.header}
\While{$\neg \phi_0$}
  \ParStart
    \State || update $M_{stategraph}$
    \ParIfStart{new states observed} 
        \State $o' \gets selectNode()$
        \State $mark(o')$ \Comment{mark it as visited}
        \State $navigateTo(o')$ using $Nav$
        \If{$o'=o \And  \neg \phi_o$}
            \State $dynamicGoal(o',\phi)$ 
        \ElsIf{$o'$ is a blocker $\And  o.isblocked$}
            \State $dynamicGoal(o', \ o'.\; \neg o'.isBlocked)$
        \EndIf
    \State \hspace{-3mm} {\bf else if}  \ there is terrain unexplored {\bf then} \ $explore()$ \label{search.alg.explore}
    \State \hspace{-3mm} {\bf else }  \ $abort()$
    \ParIfEnd
  \ParEnd
\EndWhile
\State {\bf assert} $\psi$ \label{search.alg.assert}
\EndProcedure
\end{algorithmic}
\end{algorithm}

To move forward from the agent current position, if the agent sees new states, the heuristic $selectNode()$ is invoked to select a state $o'$ for the agent to go to.
Else,
when there is no new observed state,
the agent will invoke $explore()$ to find new states.
%
%
%
If the final goal $o$ is now in $S$, it will be selected as $o'$. 
Else, an unmarked $o'$ from the set $I \cup B$ is selected.
The selected $o'$ is then marked to avoid choosing it again and causing the agent to run in an infinite loop. 
The agent then navigate from its current position to $o'$; using pathfinding over $Nav$.
Additionally, $M_{stategraph}$ is updated in parallel the whole time; it will be invoked regardless of which steps is taken; we will explain this later in Section \ref{sec.implementation}.  $Nav$ is updated by  $explore()$ in line \ref{search.alg.explore}.

If $o'$ is the final goal $o$, and it does not satisfy $\phi_o$, the
heuristic
$dynamicGoal(o',\phi_o )$ is invoked to try to change its state to $\phi_o$. Else, if $o'$ is a blocker and $o'.isBlocked$ is true, $dynamicGoal$ is invoked with $( o', \; \neg o'.isBlocked)$ as a goal, to unblock the blocker. 
Let us explain the heuristics used in $selectNode()$ and $dynamicGoal()$.

\subsubsection*{$selectNode()$} To go from one location in the level to another, we use the transition in  $M_{stateGraph}$. If we can go directly from our starting state to our goal state $o$, then life is simple. Otherwise we explore $M_{stateGraph}$ to travel through its states. This is done by selecting an intermediate state to go to.
To decide which intermediate state should be selected, we apply a policy.

 We give a higher priority to newly observed states.
 Moreover, states in $B$ have higher priority than states in $I$. Then, the distance to the approximate location of the goal, if given by developers, and the distance to the current agent position are considered. The candidate closest to the goal is preferred, and else the one closest to the agent.
If in the new observation, there is no new blocker but there is a state in $B$ which is in the agent visibility range, the state from the $B$ is selected.

\subsubsection*{$dynamicGoal(o', \eta)$} 
This procedure can be thought to deploy a goal to change $o'$ to a state satisfying $\eta$. It would try different interactables which has not been touched in this endeavour. So, we keep track of interactables that have been tried for $o'$; this is done by $mark_{o'}(i)$. 
Also note that changes on a state of an in-game object might not be immediately observable by the agent. That makes thing more complicated.

\begin{algorithm}
\footnotesize
\caption{Dynamic Goal}\label{alg:dynamicGoal}
\begin{algorithmic}[1]
\Procedure{dynamicGoal$_h$}{$o',\eta$}
   \While{$ o' $does not satisfy $\eta$}
   \State $\Delta \gets \{(i,o') \: | \; (i,o') \in P \}$ 
   \If{ $\Delta = \emptyset$}
   \State $\Delta \gets \{ j \: | \; (j,a,o') \in T,\; j {\in} I \}$ 
   \Comment{interactables nearest to $o'$}
   \EndIf
   \If{ $\Delta = \emptyset$}
    \State $\Delta \gets \{ i \: | \; i \in I, \;  \mbox{$i$ unmarked}\}$ 
    \Comment find unmarked enablers 
   \EndIf
    \If{ $\Delta = \emptyset$}
       \If{there is terrain unexplored}
      \State $explore()$
       \Else
       \State $abort()$
       \EndIf
    \Else 
    \State choose $i \in \Delta$, which is closest to the agent
    \State $mark_{o'}(i)$ \Comment{mark $i$ as touched for $o'$}
        \State $reach(i)$
        \State $interact(i)$
        \State $reach(o')$
   \EndIf
   \EndWhile
\EndProcedure
\end{algorithmic}
\end{algorithm}
\space

To minimize the effort spent to change $o'$ (to make it satisfies $\eta$), firstly, the list of pairs in $P$ will be checked to see if there is an $i$ that would affect $o'$. If not, $i$ is selected from the list $I$, if it is not empty. Interactables in $I$ having edges (transitions in $T$) to $o'$ are closer to $o'$, and are hence preferred over interactables with no edge to $o'$. If the above give multiple candidates, the $i$ closest to the agent is chosen.
To interact with $i$, the agent typically should be close enough to $i$; $reach(i)$ will try to guide the agent to $i$. Because $i$ is seen before, the agent believes that there is a path to reach it. However, on the way to $i$ it might discover that the path has become blocked, due to some previous toggling of an interactable. In this case $unstuck(i)$ is called to unblock the path.

After interacting with $i$ the agent needs to check if this actually changes $o'$ to $\eta$. However, note that $o'$ might be far from the agent. To check its state the agent needs to travel to it using $reach(o')$. The same situation as with $reach(i)$ may happen which requires invoking $unstuck(o')$. 

%
%
If all states in $I$ have been touched (no more candidates to try), $explore()$ is invoked to find a new state. The $dynamicGoal(o',\eta)$ is aborted if none of the states in $I$ can change the $o'$ state and there is no more space/states to explore. 
\subsubsection*{$Unstuck(e)$} 
Recall that this is invoked to unblock the path to a destination object $e$ that the agent tries to reach.
Note that as the agent search and explore, it also builds up the model $M$. We  use $M$ to see if it gives us a solution in the form of an interactable $i$ that would  unblock the path. The agent then interacts with $i$. Section \ref{sec.prolog} will explain how this is employed to unstuck the agent.

\subsubsection*{Example}As an example, consider a simple 'level' shown in Fig. \ref{fig.LR} taken from the game Lab Recruits. There are four buttons and three doors in this level. The player is shown at the bottom left. 
\begin{definition}\label{example.task-1}
Imagine a testing task $T_0$ where an agent has to verify that the treasure door $d_T$ is reachable and can be opened.
\end{definition}
To verify this, the agent invokes $onlineSearch(\phi_{d_T}\Rightarrow\psi,M,Nav)$,
where $\phi_{d_T} = \neg d_T.isBlocking$ and just $true$ for $\psi$.
In the algorithm, the agent first needs to find a way to reach the treasure door $d_T$.
Since the agent has a limited visibility range, it can not see the entire room. 
Imagine its visibility range is inside the red circle around the agent. The agent starts from its starting state ($c_0$). If the agent sees a state, $M_{stateGraph}$ will be updated. 
In this example, the agent can see $b_1$ and $b_2$; so they are added to $S$. As the treasure door is not in the current $S$ yet, $selectNode()$ is invoked to choose a state to move forward.
The distance from the agent position to the both of new states $b_1,b_2$ is calculated; $b_2$ is selected based on the distance. In the next step, $navigateTo(b_2)$ is called to move the agent from the current position to $b_2$. 
Then, the agent again updates $M_{stateGraph}$ as it can see new states in the new position. 

In the new observation, a blocker $d_1$ is seen. Based on the heuristic in $selectNode()$, the next $o'$ to move forward is $d_1$. Since $d_1$ is in the blocking state/closed, $dynamicGoal(d_1, \neg d_1.isBlocked)$ is invoked. The agent now switches to solve this intermediate goal which is opening $d_1$. Firstly, $P$ is checked to find a button $i$ such that $(i,d_1) \in P$.
However $P$ is still empty;
so, $T$ is checked and $b_2$, which is the nearest interactable to $d_1$, is selected and marked by $mark_{d_1}(b_2)$. After interacting with $b_2$, the agent checks the state of $d_1$. 
Suppose $d_1$ is now  open,
the goal that was set by $dynamicGoal()$ is then successfully achieved. In the current position, the agent has entered a new room. It would see more states, hence increasing the chance of reaching the treasure door. The next state to move forward is $d_2$, chosen by $selectedNode()$. Similar to $dynamicGoal(d_1, \neg d_1.isBlocked)$, the agent now tries to open $d_2$. For this, $b_3$ would selected, because it is the closest to $d_2$. The agent moves to the next room after opening $d_2$. In the new room, the agent observes $b_4$ and moves to it. 

In the current position, there is no new state that the agent can select to move. Therefore, it falls back to exploring the world. Imagine that the previous interaction with $b_3$ also closed $d_1$; the agent is then stuck in the rooms. The aforementioned $unstuck$ will be invoked to open a path out; re-toggling $b_3$ opens $d_1$ again. The agent can now explore the level; eventually it will see the treasure door.  At that time, the treasure door would be in $S$.
However, the testing task $T_0$ is not achieved yet. To verify that the treasure door can be opened, $dynamicGoal(treasure door,\neg treasuredoor.isBlocked)$ is invoked; similar to $dynamicGoal(d_1,\neg d_1.isBlocked)$.

\section{On-the-fly Model Construction} \label{sec.prolog}


Recall that the $onlineSearch$ algorithm in Section \ref{sec.online-agent-based search approach} requires a model $M$, in particular its state-graph component. In our implementation, this state-graph is represented as  Prolog facts. The implementation in Prolog gives us the flexibility to have rules for reasoning which is important in the $unstuck$ procedure used in the search algorithm. 

As it is mentioned before, $M_{stateGraph} = (S,T,\Sigma,P)$ will be gradually constructed based on what the agent observes during the search. The first three elements will be immediately updated, if new states are observed. Firstly, {\em newly observed/seen} states $N$ are added in the set of $S$, and tagged if they are interactables or blockers. E.g. if a state $s$ is a button, we register it as an $\bf interactable$, whereas a door is registered as a $\bf blocker$.
%
The next step is to update the transition set $T$. Let $s_c$ be the agent's current state. 
As the agent {\em can see} $N$  from the current state, there is thus a straight line path to navigate and reach them, with no blocker in between. 
So, transitions $s_c \rightarrow t$ and $t \rightarrow s_c$, for every $t\in N$, with the transition label 'navigateTo' are added to $T$. If a state $s_1$ is interactable, a transition from $s_1$ to itself with the transition label 'interact' will be added as well.  

To detect in which zone these states are located, or we are in a new zone, some steps need to be done. The first step is to get the current zone based on $s_{c}$. To know that newly observed states are in the current zone, one state located in the current zone is randomly selected ($s_{r}$). Then, pathfinding on $Nav$ is invoked to check if there is a path between $s_{r}$ and each one of these states when all blockers are closed. 
This is done by temporarily removing the nodes in $Nav$ that are occupied by the blockers in $S$, before invoking the pathfinder. They are put back after the zone-checking.
If there is a path, the zone of the newly observed state will be the current zone. If not, a new zone $R$ is added to $\Sigma$ with newly observed states as a member of $R$.    
%
%

Consider again the example in Fig \ref{fig.LR}. The first states that are observable by the agent at the beginning of the game are $b_1$ and $b_2$. So, they will be added to $S$.  Because both of the observed states are interactable, two transitions with the 'interact' label from each of these state to themselves are added in $T$; e.g. $ b_1 \xrightarrow{interact} b_1$. The next step is to check in which zone they are placed. Because so far $\Sigma$ is empty, a new zone with $b_1$ and $b_2$ as its member is registered $R_1 = \{ b_1,b_2 \}$ to $\Sigma$.

In the proposed approach in section \ref{sec.online-agent-based search approach}, whenever $dynamicGoal(o')$ is invoked, and solved by an interactable $i$, we record this knowledge by adding the entry $(i,o')$ to $P$. In addition, toggling $i$ may open another blocker $b$ which is on the way to $o'$. So, the pair of $(i,b \in B)$ is added to $P$ as well. 


As an example of a reasoning rule over the model, expressed in Prolog, the following states that two rooms/zones are neighbors if there is a  blocker shared by them, and hence connecting them: 
\[ neighbor(R_1,R_2) 
  \ {:}{-} 
  \left\{
  \begin{array}{l}
  R_1 \not{=} R_2, \\ 
  isBlocker(b), \\ 
  inZone(R_1,b),\ 
  inZone(R_2,b)
  \end{array} \right.
\]
%

From this rule, we can define
$roomReachability(K,R_1,R_2)$ rule as a $K$-step transitive closure of the $neighbor$-rule to  describe a condition that two non-neighboring rooms are reachable from each other because of  $K{-}1$ other rooms in between that connect them. The $unstuck$ procedure from Section \ref{sec.online-agent-based search approach} uses this rule.
Imagine the agent is in some zone $R_2$ and toggles $i$ to unblock $o'$ which is several rooms away from $R_2$. 
After toggling $i$, some blockers $d_1$ and $d_2$ in $R_2$ become closed, causing the agent to become locked in $R_2$, and hence unable to find a way back to $o'$ to check its state. Suppose $d_1$ is connected to $R_1$ which leads to $o'$, and $d_2$ is connected to $R_3$, away from $o'$. Opening one of these blockers will unstuck the agent. Using the $roomReachability$ rule allows the agent to choose the right door to open.
Note that simply re-toggling $i$ is not always an efficient way to unlock the agent, e.g. if $i$ is far from $d_1$, while there is an $i'$ next to $d_2$ which can open it. Also, if $i$ is the only interactable that can open $o'$, re-toggling it closes $o'$ again.

\section{Implementation}\label{sec.implementation}
We implement our game testing approach using iv4XR\footnote{https://github.com/iv4xr-project/aplib}, a Java multi-agent programming framework for game testing.
The framework is inspired by the popular Belief-Desire-Intent concept of agency \cite{herzig2017bdi}, where agents have their belief which represents information the agent has about its current environment and their own goals representing their desire. 
 

The framework allows tests to be programmed at a  high level, hiding underlying details such as 3D navigation and geometric reasoning. A* path finding is applied to provide an ability to auto-explore the environment/world and to auto-navigate to a game-entity, given its id (rather than its concrete position in the world) \cite{prasetya2020navigation}. 
This ability of auto navigation and exploration in-game world is used in $navigationTo(o)$ and $explore()$ mentioned in Section \ref{sec.problem-setup}. 

Recall the testing task $T_0$ from Definition \ref{example.task-1}. To solve $T_0$ the agent needs to find a right sequence of actions to reach the treasure door. Such a goal is very hard for an agent to achieve directly. It needs to be broken into subgoals to help the agent to solve the original goal, in a way such that each lowest subgoal is simple enough to be solved automatically. In iv4XR, we can define a 'goal structure' expressing such a decomposition using goal-combinators provided by the framework. More precisely, a {\em goal structure} is a tree containing basic goals as leaves and goal-combinators as nodes. Each goal at the leaves formulates certain SUT states that we want to reach, along with a so-called tactic to solve the goal. A tactic is a way to hierarchically combine basic actions using tactic-combinators.

In our approach, complicated testing tasks can be formulated purely at the goal level, without having to specify the needed tactics. The latter were implicitly provided by our implementation as part of its automation. 

\section{Experiment}\label{sec.experiment}

To evaluate our approach, we conducted a set of experiments. We use the aforementioned Lab Recruits (LR) game as a case study. It is a maze-like 3D game; a screenshot was shown in Fig. \ref{fig.LR}. We have doors as blockers, and buttons as interactables. Toggling a button toggles the state of doors that are associated to it.
LR allows new game 'levels' to be defined, which makes it suitable for experiments.
In gaming, the term {\em levels} refers to worlds or mazes that are playable in the same game.

\subsubsection*{Research Questions} 

\begin{itemize}
    \item {\bf RQ1:} {\em How effective is our online agent-based search algorithm in solving the given testing tasks?}
    \item {\bf RQ2:} {\em Can the algorithm construct an accurate model of the game under test? }
    \end{itemize}

\begin{figure}[h]
    \centering
    \includegraphics[width=\linewidth,height=5cm,scale=0.5]{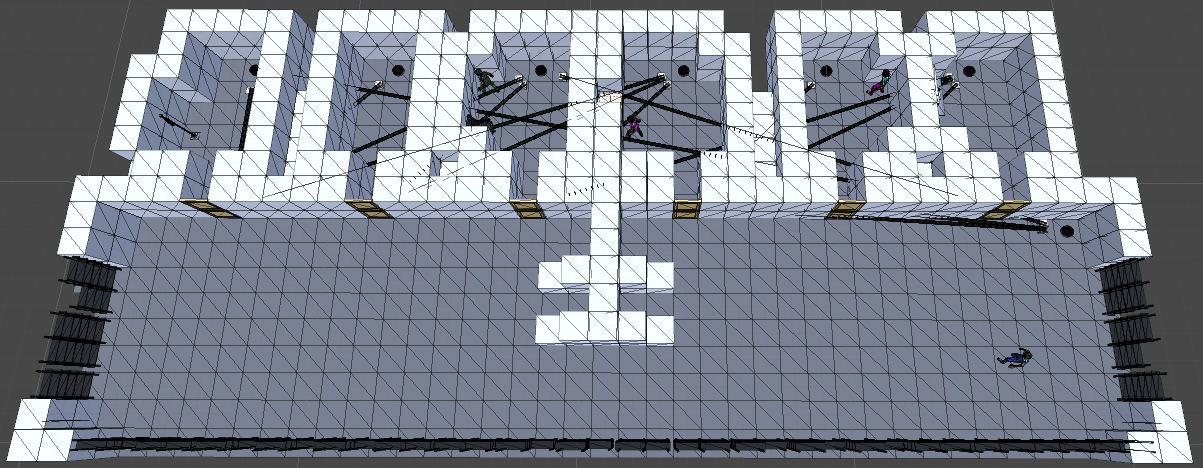}
    \caption{
   The layout of the $R7\_3\_3$ level. Lines indicate connections between buttons and doors.}
    \label{fig:A-TEST}
\end{figure}

Toward answering the research questions, we use LR levels that were used in the Student Competition in the Workshop on Automating Test case Design, Selection and Evaluation (A-TEST) in 2021\footnote{https://github.com/iv4xr-project/JLabGym/blob/master/docs/contest/contest2021.md}.
Fig. \ref{fig:A-TEST} shows the map of one of these levels ($R7\_3
\_3$). 

We applied our online search algorithm with two different setups.
 (1) The setup $Search$ uses the algorithm as in Section \ref{sec.online-agent-based search approach}. So, it exploits the on-the-fly constructed model to help  in dealing with complicated situations. For example, an interaction in the past might close a door, causing the agent to become locked in a zone. Using the model might help the agent to find a way to unlock itself; by interacting with the corresponding button to unblock the right blocker. Moreover, exploiting the model can decrease the time spent to solve a testing task, as the test agent would then know how to unblock a blocker when it faces it again. (2) In the setup $Search_{basic}$, the agent runs the same search algorithm, but it does not have access to the constructed model; it can not thus exploit the model.

A $Random$ test algorithm is also applied to serve as a baseline. This $Random$ repeatedly alternates between exploring a given level to discover game objects, and randomly choosing a pair of button and door; it then toggles the button to find out if it opens the door. If so the connection is recorded. This is repeated until its budget runs out; we set this budget to be $1.2
T$ where $T$ is the time used by $Search$ to solve the same testing task. For each level $Random$ is run ten times.

The testing tasks posed to all levels is to verify that a chosen closed door $o$ is reachable and can be opened. This chosen $o$ is always a door that is important for completing the level, and whose reachability is non-trivial.  
This corresponds to the $\phi$-part in (\ref{eq.testingtask}). The assertion part $\psi$ is less important for this study, so it is just $true$.
To reach $o$, a sequence of actions is required to be done by the agent. This sequence is {\em not} known upfront; only the id of $o$ is given by developers to the agent; the agent should find the solution (the aforementioned sequence) by itself.

\subsubsection*{Levels}
The  A-TEST levels provide a range of size and complexity to test the algorithm. 
For a blocker $o \in O$, let $\mu(o)$ be the number of interactables that can toggle $o$ (in LR, $o$ would be a door and $\mu(o)$ is the number of buttons connected to this door). For a game level, the $\mu$ of this level is the {\bf greatest} $\mu(o)$ over all blockers in the level. 
Similarly, for an interactable $i \in O$, $\nu(i)$ is the number of blockers that $i$ can toggle. The $\nu$ of a game level is defined as the greatest $\nu(i)$.
Setups with $\mu$ and $\nu$ higher than 1 are complicated to solve.

 The left part of Table \ref{tab:level.features} shows the features of the LR levels used in our experiments. 
 These levels have different difficulty for the agent toward solving the corresponding testing task.
 For example, some have $\nu,\mu{>}1$. 
  Some have doors which are initially open (the column $init$), which makes searching for a solution 
  even more complicated, as during the search the agent needs to try different buttons, and one of them might actually close a door that was initially open.

 \begin{table}
  \caption{Levels' features. Each contributes to their complexity. $R$, $B$, $D$ indicate the number of rooms, buttons and doors in each level; 
  $init$ specifies the number of doors which are initially open. The last three columns will be explained later.
 }\label{tab:level.features}
  {\scriptsize
  \begin{tabular}{ccccccc|ccc}
    \toprule
    level & R & B & D & $\nu$ & $\mu$ & init & $Search$ & $Search_{base}$ & $Random$\\
    \midrule
    $R3\_1\_1\_H$ & 3 & 6 & 4 & 1 &  1 &  0 & 1 & 0& 0.3 \\
    $R4\_1\_1$ & 5 & 8 & 6 &  1 &  1 &  0 & 1 & 1& 0.1\\
    $R4\_1\_1\_M$ & 4 & 8 & 6 &  1 &  1 &  0 & 1 & 0& 0.7\\
    $R5\_2\_2\_M$ & 5 & 7 & 4 &  2 & 2  &  0 & 1 & 0& 0.1\\
    $R7\_2\_2$ & 7 & 7 & 6 &  2 & 2  &  0 & 1 & 1& 0.6 \\
    $R4\_2\_2$ & 5 & 8 & 6 &  2 & 2  &  ${>}0$ & 1 & 1& 0.9\\
    $R4\_2\_2\_M$ & 5 & 7 & 4 &  2 & 2  &  ${>}0$ & 1 & 0&0.4\\
    $R7\_3\_3$ & 7 & 7 & 6 &  3 & 4  &  0 & 1 & 1 & 0.9\\
    \bottomrule
  \end{tabular}
  }
  \vspace{-0.5mm}
\end{table}

\subsection{Results}

\subsubsection{Evaluating the Ability to Find Solutions}\


{\bf RQ1:} How effective is our online agent-based search algorithm in solving the given testing tasks?

To evaluate this, all levels in Table \ref{tab:level.features} are tested by an agent. 
A testing task to open a door called the treasure door is given to the agent. 
As the ground truth, 
the tasks are solvable and the corresponding test should pass. The strength of our algorithm in solving non-trivial tasks is assessed by the number of tests that pass.


The last three columns in Table \ref{tab:level.features} show the results for $Search$, $Search_{base}$, and $Random$; $1$ means the corresponding testing task is passed and $0$ means it fails. For $Random$, a value $p$ means that it gives a pass verdict with probability $p$, sampled over 10 runs.
$Search$ successfully solves the testing tasks on all levels, including the more complex levels such as $R4\_2\_2\_M$. 
In contrast,
$Search_{base}$ is not always successful (Table \ref{tab:level.features}), implying that exploiting the model is essential for solving the testing tasks.
After looking at the failure cases, we conclude  that not only the functional relation between the objects, but also the physical layout of the level plays a role to solve a testing task without model exploitation.  
$Random$ solves the testing tasks with about 0.5 probability. Note that this also means that it has 0.5 probability to give a false positive (falsely reporting a bug) which makes it unfit for actual use.

\subsubsection{Evaluating the Constructed Model} \ 

{{\bf RQ2:}} Can our online agent-based search algorithm construct an accurate on-the-fly model of the game under test?

To verify if the constructed model is accurate, we compare the model against the actual level definition.  
In the constructed model, we have the information about the number of zones, interactables, blockers, and the $P$ component. Also, the information about the existing objects in each room can be found in the generated model. 
The $S$ (states), $\Sigma$ (zones), and $P$ components of the model are checked manually. 

Table \ref{table.model.accuracy} shows the results of $Search$ and the $Random$ algorithms; $Search_{base}$ is not included as Table \ref{tab:level.features} already showed that it is inferior to $Search$. 
We can see that all but one button-door connections that $Search$ registered in the model are correct in all levels. 
In contrast, $Random$ is quite obviously more prone to incorrectly registering connections.
The results of $Search$ indicate how reliable the on-the-fly constructed table $P$ is when the agent exploits the model to solve testing tasks. 
Also note that despite the inaccuracy, all testing tasks are still solved (Table \ref{tab:level.features}).
Some of the mistakes in $P$ are  acceptable as the agent can not immediately observe the effect of toggling an interactable if the corresponding blocker is not in the agent visibility range.

Table \ref{table.model.accuracy} also shows that most, but indeed not all, objects in each levels are recorded by $Search$ in the model it constructs. 
Keep in mind that these data are recorded only by giving one testing task (reaching and opening the treasure door) to the agent. 
We can also see that the number of connections registered by 
$Search$ and $Random$
is often almost the same,
while the latter is given 20\% more time budget. 
Finding all objects and connections is not necessary, as long the task is solved. 
However, if desired, we can
apply different testing tasks to  obtain a more complete model, e.g.  to make sure that all interesting objects are registered.

To evaluate the efficiency of our algorithm, we measure the total time to solve each testing task. We also measure the time spent for purely exploring the level (when the agent does $explore()$) and the number of blockers the agent tried to open until it solves the task. Table \ref{table.resultWithModel} shows that the run-time of $Search$ ranges between one to four minutes. Note that the agent needs to travel between various locations, e.g. to check them. Such travel simply takes time. 
Table \ref{table.resultWithModel}  also shows that the time spent exploring the game world ranges between about 15\% - 30\% of the total time. The remaining time is basically spent on actually solving the testing task; the more proportion of time spent for this is the better.


\begin{table}
  \caption{The accuracy of the constructed model for each level. The column $B/B$ shows the number of  registered buttons versus the number of all available buttons in each level. 
  Similarly we have $R/R$ and $D/D$ for rooms and doors. $C/C$ and $R(C/C)$ show the number of button-door connections registered in the $P$ component by $Search$ and $Random$ respectively.
  $W_{C}$ and  $R(W_{c})$  are the number of these recorded connections, by $Search$ and $Random$  respectively, which are {\em wrong}. For $R(C/C)$ and $R(W_{c})$ the number is the average over ten runs. $W_b$ and $W_d$ are the number of buttons and doors which are registered, by $Search$, in wrong rooms.}
  \label{table.model.accuracy}
  {\scriptsize
  \begin{tabular}{cccccccccl}
    \toprule
    level& R/R & B/B & D/D & C/C & R(C/C) & W$_{c}$ & R(W$_{c}$)& W$_{b}$ & W$_{d}$\\
    \midrule
    $R3\_1\_1\_H$  & $2/3$ & $5/6$ & $2/4$ & $2/4$ &$2.3/4$& 0 &0& 0 & 0 \\
    $R4\_1\_1$ & $4/5$ & $8/8$ & $5/6$ & $5/6$ &$4.8/6$& 0&0 & 0 & 0\\
    $R4\_1\_1\_M$  & $3/4$ & $8/8$ & $5/6$ & $5/6$ &$4.5/5$& 1&0 & 0 & 0\\
    $R5\_2\_2\_M$  & $5/5$ & $5/7$ & $4/4$ & $5/6$ &$3.1/6$& 0&0.1 & 2 &1 \\
    $R7\_2\_2$  & $5/7$ & $4/7$ & $6/6$ & $7/11$ &$8.3/11$& 0& 1.9 & 0 & 0\\
    $R4\_2\_2$  & $4/5$ & $7/8$ & $5/6$ & $1/9$ &$5.2/9$& 0 &0& 0 & 0\\
    $R4\_2\_2\_M$  & $5/5$ & $5/7$ & $4/4$ & $5/7$ &$4.3/7$& 0 & 0.5 & 0 & 0\\
    $R7\_3\_3$ & $4/7$ & $3/7$ & $6/7$ & $7/16$ &$14.9/16$& 0 &2.7& 0 & 1\\
    \bottomrule
  \end{tabular}
  }
  \vspace{-0.5mm}
\end{table}

\begin{table}
  \caption{The performance of the algorithm on experiment's levels;
  $time$ and $exporation$ show the total run time and the time spent purely on exploration in $Search$ algorithm.
 }\label{table.resultWithModel}
 {\small
  \begin{tabular}{cccccl}
    \toprule
    level &  tried doors &  time (s) & exploration (s) \\
    \midrule
    $LR3\_1\_1\_H$ & 3 & 68 &  14\%  \\
    $R4\_1\_1$ &  5 &  84 &  22\% \\
    $R4\_1\_1\_M$ & 5 &  139 &  17\%  \\
    $R5\_2\_2\_M$ &  6 &  140  & 19\% \\
    $R7\_2\_2$  & 4 &  146 & 28\%  \\
    $R4\_2\_2$  & 1 &  60 & 33\%  \\
    $R4\_2\_2\_M$  & 4 & 144 & 28\% \\
    $R7\_3\_3$ & 6&  254 & 22\% &\\
    \bottomrule
  \end{tabular}
}  
  \vspace{-5mm}
\end{table}

\vspace{-3mm}
\section{Related work}\label{sec.relatedwork}

Recently, testing has become an increasingly important instrument for improving  the quality of computer games. 
Research  has provided various methods \cite{ostrowski2013automated,iftikhar2015automated} towards automated game testing, but they still require substantial manual work, e.g. to prepare models \cite{iftikhar2015automated} or to redesign and re-record test sequences when the game is changed. Hence, researchers  have been investigating ways to combine automated testing and the application of techniques from machine learning \cite{zheng2019wuji,ariyurek2019automated,zarembo2019analysis} in the context of game testing. E.g. Pfau et al. \cite{pfau2017automated} developed ICARUS to test and detect bugs in an adventure game. 
Using an artificial agents to create player personas and letting them evolve through playing is another recent approach used in automated game testing \cite{mugrai2019automated,ariyurek2019automated}. To approximate different play styles, Mugrai et al. \cite{mugrai2019automated} developed different procedural personas through the utility function for a Monte Carlo Tree Search (MCTS) agent. Similarly, Holmgard et al. \cite{holmgaard2018automated} described a method for generative player modeling through procedural personas and its application to the automatic game testing. Agents are used to help playtest games as well \cite{borovikov2019towards, silva2018exploring,stahlke2019artificial}. Zhao et al. tried to build agents with human-like behaviour, aiming to help with game evaluation and balancing \cite{zhao2020winning}.  
However, all such types of AI also require much training, which could make them impractical to be deployed during the development time where SUT would undergo frequent changes.




Model-based testing \cite{utting2012taxonomy} is a well known automated testing approach which has been used in various studies \cite{akpinar2020web,panizo2020model}.
However, its application
in computer games has not been much studied. Some that we can mention is e.g the work of 
Iftikhar et al. \cite{iftikhar2015automated} that used a UML-based model to support automated system-level game testing of platform games.
Ariyurek et al. \cite{ariyurek2019automated} use a scenario graph, which is essentially an FSM, for generating asbtract test sequences. A reinforcement learning (RL) and MCTS 
agent is used to find a concrete sequence of actions that realizes each abstract test sequence. 
A more recent study is done by Ferdous et al. \cite{ferdous2021search} that proposed an EFSM model for modelling game behaviour and combined it with search-based testing for test generation. 
Generating and executing tests are automated. 
However, models often have to be manually constructed. which requires a lot of efforts. 

There are techniques that enables a computer to construct models, e.g. by 'inferring' them from execution traces as in \cite{lorenzoli2008automatic,lo2009automatic,foster2019incorporating}. In \cite{lo2009automatic}, Lo et.al use 
a two staged inference: first a set of simple temporal properties are statistically mined from the trace, then they are used to guide the construction of a generalizing FSM.
Lorenzoli et al. \cite{lorenzoli2008automatic} present a dynamic analysis technique using Daikon 
to automatically generate an EFSM model of the system under test from the interaction traces that also contain  data values. 
The models inferred by these approaches are only applicable to trace with specific characteristic, and depends on the quality of execution samples used to produced them. 

Although these approaches are automated, they use data traces to capture the EFSM that limit its effect on modeling modern games with high-level interactivity.
On-the-fly model construction, such as used in our algorithm, is very different from trace-based model inference.
The latter requires multiple executions, whereas in an on-the-fly construction we only have one execution, though on the other hand the test agent has control on how the execution proceeds. 


\section{Conclusion} \label{sec.concl}
This paper focused on the challenges of automated testing on modern computer games.  
We proposed an online search algorithm on top of the agent-based testing framework with on-the-fly model construction. Having an on-line search means a full pre-constructed model of the game under test is not required. 
The online 
algorithm can deal with dynamic obstacles that can block the agent access to other objects. In this study we do not consider hazard and mobile objects and we restrict ourselves to toggling switches; this is done in a separate study outside the scope of this paper.
Based on the applied heuristics, an agent explores the 3D game world to solve the given testing task and unblocks the obstacles in its way. To aid the search, an EFSM model is defined to capture only general properties of the game; the remaining part of the model is constructed on the fly, which is then exploited to solve the testing task.

To evaluate our approach, we conducted a set of experiments. We used benchmarking levels that have different difficulty. It was observed that the agent can successfully solve the given testing tasks at all levels using the online search algorithm and exploiting the constructed model. The constructed model is also verified by comparing the result of the data set registered in the constructed model with the actual data defined in each level. The results show that the generated model is mostly correct and almost complete. 

In the future, we would like to study how to improve the accuracy of the constructed model to have a full model of the game under test. Also, we would like to investigate how to exploit the model in a mixed online and offline search.

\bibliographystyle{ACM-Reference-Format}
\bibliography{Bibliographies}
\end{document}